\documentclass[pra,aps,twocolumn,amsmath,amssymb,amsfonts,showpacs]{revtex4}

\usepackage{graphicx}
\usepackage{bbm,epsfig}

\newcommand{\be}{\begin{equation}}
\newcommand{\ee}{\end{equation}}
\newcommand{\ba}{\begin{array}}
\newcommand{\ea}{\end{array}}
\newcommand{\bea}{\begin{eqnarray}}
\newcommand{\eea}{\end{eqnarray}}
\newcommand{\tr}{\mbox{Tr}}                                        
                                        
\newcommand{\bra}[1]{\ensuremath{\langle #1 |}}
\newcommand{\ket}[1]{\ensuremath{| #1 \rangle}}

\newcommand{\ad}{{a^{\dagger}}}

\begin{document}

\title{Cavity driven by a single photon: conditional dynamics and non-linear phase shift}

\author{A. R. R. Carvalho$^{1}$, M. R. Hush$^{2}$, M. R. James$^{2,3}$}
\affiliation{$^{1}$ARC Centre for Quantum Computation and Communication Technology, Department of Quantum Science, Research School of Physics and Engineering,
The Australian National University, Canberra, ACT 0200 Australia}
\affiliation{$^{2}$Research School of Engineering, The Australian National University, Canberra, ACT 0200, Australia}
\affiliation{$^{3}$ARC Centre for Quantum Computation and Communication Technology, The Australian National University, ACT 0200, Australia}

\begin{abstract}
We apply the stochastic master equations (quantum filter) derived by Gough {\it et al.}~\cite{Gough:2011a,Gough:2011b,Gough:2011} to a system consisting of a cavity driven by a multimode single photon field. In particular, we analyse the conditional dynamics for the problem of cross phase modulation in a doubly resonant cavity. Through the exact integration of the stochastic equations, our results reveal features of the problem unavailable from previous models.
\end{abstract}
\pacs{42.50.Lc,  03.65.Yz, 42.50Dv}
\maketitle

\section{Introduction}

Single photons constitute an important element for a variety of quantum technological applications. Sometimes referred as a ``flying qubit", propagating single photon wavepackets can serve, for example, as the qubits used for linear-optical quantum computation~\cite{Knill:2001}, or as carriers of quantum information between nodes in a quantum network~\cite{Cirac:1997,Cirac:1998,Kimble:2008}. 

These possibilities have triggered an enormous experimental effort toward the production of single photons on demand~\cite{Grangier:2004}. Theoretically, the focus has been on the development of a proper description of the continuous-time multimode single photon field and its interaction with other quantum systems, such as atoms or cavities. In Ref.~\cite{Gheri:1998}, for example, Gheri {\it et al.} introduced a formalism to describe the dynamics of a quantum optical system driven by a single photon field in terms of a generalised master equation. A proper treatment for the multimode single photon field has also been taken into account in some photonic quantum gates~\cite{Shapiro:2007,Munro:2010} and also in the analysis of the loading probability in trapped-atom quantum memories~\cite{Razavi:2007}. 

More recently, Gough {\it et al.}~\cite{Gough:2011a,Gough:2011b,Gough:2011}  developed a stochastic master equation (SME)~\cite{Carmichael:1993, Wiseman:2010}, or quantum filter~\cite{Belavkin:1987,Belavkin:1992,Barchielli:1991} to describe the situation where a quantum system is continuously monitored and driven by a variety of non-classical states, including single photon fields. This description is important not only to analyse the conditional dynamics of systems undergoing continuous measurement but is also an essential step to model measurement-based feedback. 

Quantum feedback control has been recently used in a range of problems from the  preparation and protection of quantum states~\cite{Carvalho:2007,Sayrin:2011,Stevenson:2011a} to the stabilisation of Bose Einstein condensates~\cite{Wilson:2007,Szigeti:2009}. In the same way that feedback and other control techniques became ubiquitous in classical technological applications, we expect that they will play a similar role in future quantum enabled technologies. In a network of quantum systems, for example, where one could imagine that some of the channels connecting nodes will be fed back to the system while others will be measured, it becomes essential to provide a physical description of the conditional dynamics of such processes. 

In this paper we apply the formalism developed in~\cite{Gough:2011a,Gough:2011b,Gough:2011} to describe the conditional dynamics of the cross phase modulation in a doubly resonant cavity. In this problem, a single photon pulse interacts with a coherent field through a cavity nonlinearity inducing a phase shift in the coherent mode. Our goal here is to investigate how a proper stochastic model and its solution can bring new information about the conditional phase shift process. The paper is organised as follows: In Section~\ref{sec:model} we present the stochastic Schr\"odinger equations (SSE) corresponding to photodetection and homodyne measurement of the single photon channel. This is an alternative to the SME derived in~\cite{Gough:2011a,Gough:2011b,Gough:2011} and an explicit construction of the master equation unravelling proposed in~\cite{Gheri:1998}. In Section~\ref{sec:cav} we analyse the simple model of a driven single mode cavity. The simplicity of the system allows us to develop intuition about the driving and the monitoring. In Section~\ref{sec:phase} we discuss the model~\cite{Munro:2010} of a doubly resonant cavity that is driven by a single photon pulse and a coherent field. We analyse the problem of the non-linear cross phase shift imprinted by the single photon on the coherent field conditioned on the measurement of the single photon mode. Finally, in Sec.~\ref{sec:conc} we conclude and discuss future perspectives.

\section{The model: conditional evolution}
\label{sec:model}
\begin{figure}[h]
\includegraphics[width=0.7\linewidth]{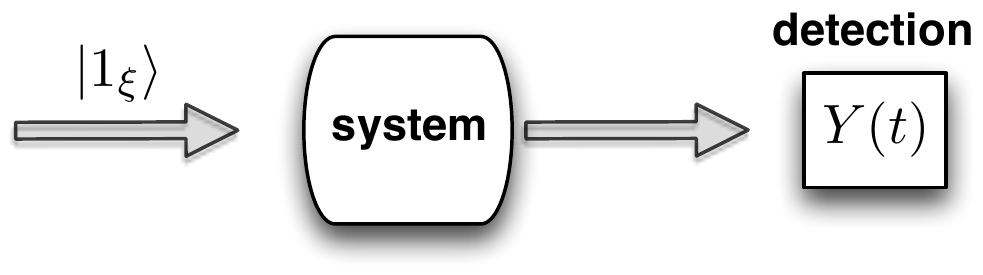}
\caption{Schematics of a quantum system probed by a single photon field. The dynamics of the system will be conditioned on the measurement outcomes $Y(t)$.}
\label{fig:probe}
\end{figure}

We start this section with a brief summary of some of the results from~\cite{Gough:2011a,Gough:2011b,Gough:2011} to establish the notation and prepare the ground to present the SSE for a single photon channel. The general scenario we want to describe is the one shown in Fig.~\ref{fig:probe}, where a continuously monitored system is probed by a single photon field. The state of this multimode field is given by $\ket{1_{\xi}}=\int_{0}^{\infty} dt \xi(t) \hat b^{\dagger}(t) \ket{0}$~\cite{Loudon:1973}, where $\int_0^{\infty}dt  |\xi(t)|^2=1$. 

Before moving into the monitored evolution, it is instructive to look at the unconditional dynamics given by the master equation. Let $\rho$ be the density matrix for the joint state of the field and the system. Now, in order to describe the dynamics of the system itself one needs to trace over the field variables. When the field is in a single photon state, the relevant unconditional system density operator is given by $\rho_{11}(t)=\bra{1_{\xi}} \rho \ket{1_{\xi}}$. The master equation for the system derived in~\cite{Gheri:1998,Gough:2011a,Gough:2011b,Gough:2011} is given by
\begin{subequations}
\label{eq:me}
\begin{align}
\frac{d \rho_{11}}{dt}&= {\cal L}\rho_{11}(t) + [\rho_{01}(t),L^{\dagger}] \xi(t) +[L,\rho_{10}(t)]\xi^*(t) , \\
\frac{d\rho_{01}}{dt}&= {\cal L}\rho_{01}(t) + [L,\rho_{00}(t)] \xi^*(t) , \\
\frac{d\rho_{10}}{dt}&= {\cal L}\rho_{10}(t) + [\rho_{00},L^{\dagger}]\xi(t), \\
\frac{d\rho_{00}}{dt}&= {\cal L}\rho_{00}(t),
\end{align}
\end{subequations}
where the Liouvillian term  ${\cal L}\rho(t)=-i[H,\rho]+{\cal D}[ L]\rho$ contains the Hamiltonian and the decoherence part in the usual Lindblad form ${\cal D}[L]\rho=L\rho L^{\dagger} -1/2(L^{\dagger} L \rho +\rho L^{\dagger}L)$. An important consequence of the field being in a single photon state is that the dynamical equation for the system operator $\rho_{11}(t)$ depends also on the operators $\rho_{00}$, $\rho_{01}$, and $\rho_{10}$. This is in contrast with the situation where the field is in a Gaussian state and there is only one equation for the system operator $\rho_s \equiv \tr_{\rm field} [\rho]$. Note, however, that $\rho_{11}(t)$ is the only physically relevant operator since the expectation value of any system operator $X_s$ can be calculated directly from it using $\langle X_s \rangle(t)=\tr{[X_s \rho_{11}(t)]}$. 

Now we want to describe the dynamics of the monitored system. From the general filtering equations derived in~\cite{Gough:2011a,Gough:2011b,Gough:2011}, we can write the conditional master equation for the system. In the case of homodyne monitoring they are 
\begin{widetext}
\begin{subequations}
\label{eq:smehom}
\begin{align}
d\rho_{11}(t)&=\left[ {\cal L}\rho_{11}(t) + [\rho_{01}(t),L^{\dagger}] \xi(t) +[L,\rho_{10}(t)]\xi^*(t)  \right] dt \nonumber \\
&+\left[ L\rho_{11}(t)+ \rho_{11}(t) L^{\dagger}  +\rho_{10}(t) \xi^*(t) +\rho_{01}(t)\xi(t) -K_t \rho_{11}(t) \right] dW(t), \quad \quad \\
\nonumber \\
d\rho_{01}(t)&=\left[ {\cal L}\rho_{01}(t) + [L,\rho_{00}(t)] \xi^*(t) \right]dt+\left[L\rho_{01}(t)+\rho_{01}(t) L^{\dagger}  +\rho_{00}(t) \xi^*(t)  -K_t \rho_{01}(t) \right] dW(t), \quad \quad \\
\nonumber \\
d\rho_{10}(t)&=\left[ {\cal L}\rho_{10}(t) + [\rho_{00},L^{\dagger}]\xi(t) \right]dt +\left[L\rho_{10}(t)+\rho_{10}(t) L^{\dagger}  +\rho_{00}(t) \xi(t)  -K_t \rho_{10}(t) \right] dW(t),  \quad \quad \\
\nonumber \\
d\rho_{00}(t)&= {\cal L}\rho_{00}(t) dt+\left[L\rho_{00}(t)+\rho_{00}(t) L^{\dagger}   -K_t \rho_{00}(t) \right] dW(t),
\end{align}
\end{subequations}
\end{widetext}
with $dW(t)$ a Wiener process obeying $dW(t)=dY(t)-K_t dt$ and $K_t=\tr\left[(L+L^{\dagger})\rho_{11} +\rho_{01}\xi(t) +\rho_{10} \xi^*(t)\right]$. For the case of a photocounting monitoring the equations become~\cite{Gough:2011}
\begin{widetext}
\begin{subequations}
\label{eq:smepd}
\bea
d\rho_{11}(t)&=&\left[ {\cal L}\rho_{11}(t) + [\rho_{01}(t),L^{\dagger}] \xi(t) +[L,\rho_{10}(t)]\xi^*(t)  \right] dt \nonumber \\
&+&\left[\nu_t^{-1}\left( L\rho_{11}(t)L^{\dagger}  +L\rho_{10}(t) \xi^*(t) +\rho_{01}(t)L^{\dagger}\xi(t) +\rho_{00}\vert\xi(t)\vert^2\right)-\rho_{11}(t) \right] dN(t), \quad \quad \\
\nonumber \\
d\rho_{01}(t)&=&\left[ {\cal L}\rho_{01}(t) + [L,\rho_{00}(t)] \xi^*(t) \right]dt +\left[ \nu_t^{-1}\left( L\rho_{01}(t)L^{\dagger} +L\rho_{00}(t) \xi^*(t) \right) -\rho_{01}(t) \right] dN(t), \quad \quad \\
\nonumber \\
d\rho_{10}(t)&=&\left[ {\cal L}\rho_{10}(t) + [\rho_{00},L^{\dagger}]\xi(t) \right]dt +\left[ \nu_t^{-1}\left( L\rho_{10}(t)L^{\dagger} +\rho_{00}(t)L^{\dagger} \xi(t) \right)  -\rho_{10}(t) \right] dN(t), \quad \quad \\
\nonumber \\
d\rho_{00}(t)&=& {\cal L}\rho_{00}(t) dt +\left[\nu_t^{-1}  L \rho_{00}(t)L^{\dagger} -\rho_{00}(t) \right] dN(t),
\eea
\end{subequations}
\end{widetext}
where $dN(t)$ is a compensated Poisson process obeying $dN(t)=dY-\nu_t dt$, with $\nu_t=\tr[L^{\dagger} L \rho_{11} +L\rho_{10}\xi^*(t) +L^{\dagger} \rho_{01} \xi(t) +\rho_{00} |\xi(t)|^2]$. Note that by taking the classical ensemble averages over the noisy processes we obtain back the master equations Eqs.(\ref{eq:me}).

A crucial aspect of the single photon master equations is their non-Markovian behaviour. This is important as SSE unravellings of the master equation are only guaranteed to exist if the later is Markovian. The solution to find the SSE corresponding to Eqs.(\ref{eq:smehom}) and (\ref{eq:smepd}) is to embed the system in a larger Hilbert space that contains an ancilla which generates the input field state. Such Markovian embedding has been suggested in~\cite{Gheri:1998} for the case of a single photon field and derived in~\cite{Gough:2011b,Gough:2011} for other non-classical input fields in the context of quantum filtering theory. 

Finding this ancilla for general input states is a hard task but can be simple for a single photon field. Gheri {\it et al.}~\cite{Gheri:1998} proposed a cavity initially with one photon as the auxiliary emitter while Gough {\it et al.}~\cite{Gough:2011b,Gough:2011} suggested a two-level system initially in the excited state. Using the latter as the choice for our ancilla and using a cascade approach we can write a master equation for the total system (cavity + ancilla) as 
\bea
\label{eq:fullme}
\frac{d\varrho}{dt}=-i\left[H_T,\varrho\right]+{\cal D}\left[L_T\right]\varrho,
\eea
where 
\bea
\label{eq:jump}
L_T=a+\frac{\xi(t)}{\sqrt{w(t)}}\sigma_-,
\eea
and 
\bea
H_T=H+\frac{i}{2\sqrt{w(t)}}(\xi^*(t)\sigma_+ a  -\xi(t)  \ad \sigma_-),
\eea
where $\sigma_+=(\sigma_-)^{\dagger}=\ket{e}\bra{g}$ and $\ket{e}$($\ket{g}$) represent the excited(ground) state of the two-level ancilla. As already suggested in~\cite{Gheri:1998}, from here it is easy to write quantum trajectory unravellings as Eq.(\ref{eq:fullme}) is Markovian and in the Lindblad form.  Explicitly, the stochastic Schr\"odinger equations for homodyne and photocounting are, respectively
\bea
\label{eq:ssehom}
d\ket{\Psi}&=&\left(-i H_T+ \langle L_T^{\dagger}\rangle L_T -\frac{L_T^{\dagger} L_T}{2} -\frac{ \langle L_T^{\dagger}\rangle\langle L_T\rangle}{2}\right)\ket{\Psi}dt \nonumber \\&+& \left(L_T- \langle L_T\rangle \right)\ket{\Psi} dW,
\eea
and
\bea
\label{eq:ssepd}
d\ket{\Psi}&=&\left(-i H_T +\frac{\langle L_T^{\dagger} L_T \rangle}{2} -\frac{L_T^{\dagger} L_T }{2}\right) \ket{\Psi}dt \nonumber \\&+& \left( \frac{L_T}{\sqrt{\langle L_T^{\dagger}L_T \rangle}}-1\right) \ket{\Psi} dN,
\eea
with the point process $dN$ obeying $dN^2=dN$ and ${\mathbb E}[dN(t)]=\langle L_T^{\dagger} L_T \rangle dt$. The connection between Eqs.(\ref{eq:ssehom}) and (\ref{eq:smehom}) and Eqs.(\ref{eq:smepd}) and (\ref{eq:ssepd}) can be established using the correspondences~\cite{Gheri:1998,Gough:2011b,Gough:2011}  
\bea
\label{eq:corresp}
\rho_{00}&=&\frac{\varrho_{ee}}{w(t)},\\
\rho_{01}&=&\frac{\varrho_{eg}}{\sqrt{w(t)}},\\
\rho_{10}&=&\frac{\varrho_{ge}}{\sqrt{w(t)}},\\
\rho_{11}&=&\varrho_{ee}+\varrho_{gg},
\eea
with $w(t) =\int_t^{\infty} |\xi(s)|^2 ds$ and $\varrho_	{ij}=\bra{i}\Psi\rangle \langle \Psi\ket{j}$. 

The advantage of the SMEs (\ref{eq:smehom}) and (\ref{eq:smepd}) is that they describe the conditional dynamics in terms of system operators only. The price paid is the necessity of extra coupled equations for the auxiliary quantities $\rho_{00}$, $\rho_{01}$ and $\rho_{10}$. On the other hand, the description in terms of pure states is given by a single stochastic equation ((\ref{eq:ssehom}) or (\ref{eq:ssepd}), depending on the choice of measurement process), but requires extending the system with the addition of an auxiliary 2-level system. Numerically, the SSEs have the advantage of simulating at the state vector level requiring $2\times N$ levels (with $N$ the dimension of the system of interest) instead of the $4\times N^2$ for the SMEs.
\\

\section{Simple example: Cavity driven by a single photon}
\label{sec:cav}

We will now consider the situation where the system of interest is a single mode cavity with decay rate $\kappa$. In this case, the conditional dynamics of the driven cavity is given by the models described in the previous section with $L=\sqrt{\kappa} a$, where $a$ is the annihilation operator for the cavity field. To start with, let's analyse the unconditional evolution and look at the dynamics of the mean number of photons in the cavity assuming that it starts in the vacuum. Averaging over the noise processes in Eq.~(\ref{eq:smehom}) or (\ref{eq:smepd}) we can write the system of equations necessary to describe the evolution of $\langle n\rangle_{11}\equiv \tr[\rho_{11} \ad a]$ as
\bea
\label{eq:avgN}
\frac{d \langle n\rangle_{11}}{dt}&=&\left[-\kappa \langle n\rangle_{11} -\sqrt{\kappa}( \langle a\rangle_{01} \xi^*(t) -\langle a\rangle^*_{01}\xi(t)) \right] , \\
\label{eq:avgA}
\frac{d\langle a\rangle_{01}}{dt}&=&\left[ -\frac{\kappa}{2} \langle a\rangle_{01} -\sqrt{\kappa} \xi(t) \right],
\eea
where for an arbitrary system operator $c$ we define $\langle c \rangle_{ij}= \tr[(\rho_{ij})^{\dagger} c]$ and we have at the initial time $\langle a\rangle_{ij}(0)=\langle n\rangle_{ij}(0)=0$. For the single photon shape we choose $\xi(t)=\sqrt{\gamma}e^{-\gamma(t-t_0)/2}\Theta(t-t_0)$, with $\Theta(t-t_0)$ being the Heaviside step function, which corresponds to a single photon emitted by a two level atom with decay rate $\gamma$. For this case, the exact solution 
$n_{11}(t)=4 \gamma  \kappa  e^{\gamma  (-t)} \left(e^{\frac{1}{2} (\gamma -\kappa ) (t-t_0)}-1\right)^2 \Theta (t-t_0)/(\gamma -\kappa )^2$ is shown in Fig.{\ref{fig:avgN}-a for different values of $\gamma/\kappa$. The larger the ratio $\gamma/\kappa$, the faster the photon enters the cavity but the maximum number of photons in the cavity is obtained when the single photon and cavity rates coincide, i.e. for  $\gamma/\kappa=1$. Note that Eqs.(\ref{eq:avgN}) and (\ref{eq:avgA})  are identical to the ones corresponding to the situation of a cavity driven by a coherent field with amplitude $\epsilon(t)=i\sqrt{\kappa} \xi(t)$. This means that just by looking at the average number of photons in the cavity one cannot distinguish between a single photon or an equivalent coherent driving.

Obviously this does not hold true for higher order moments as the distributions for the field inside the cavity are different in each case: for the coherent driving the vacuum is displaced but the state remains coherent while the single photon driving mixes a single photon state with the vacuum component of the field as shown by the Wigner function plot in Figure~\ref{fig:wigner}. %\footnote{See Supplemental Material at http://link.aps.org/ supplementalxxxxx for an animation with the conditional and unconditional time evolution of the Wigner function for single photon and coherent driving}. 

\begin{figure}[htb]
\includegraphics[width=8cm]{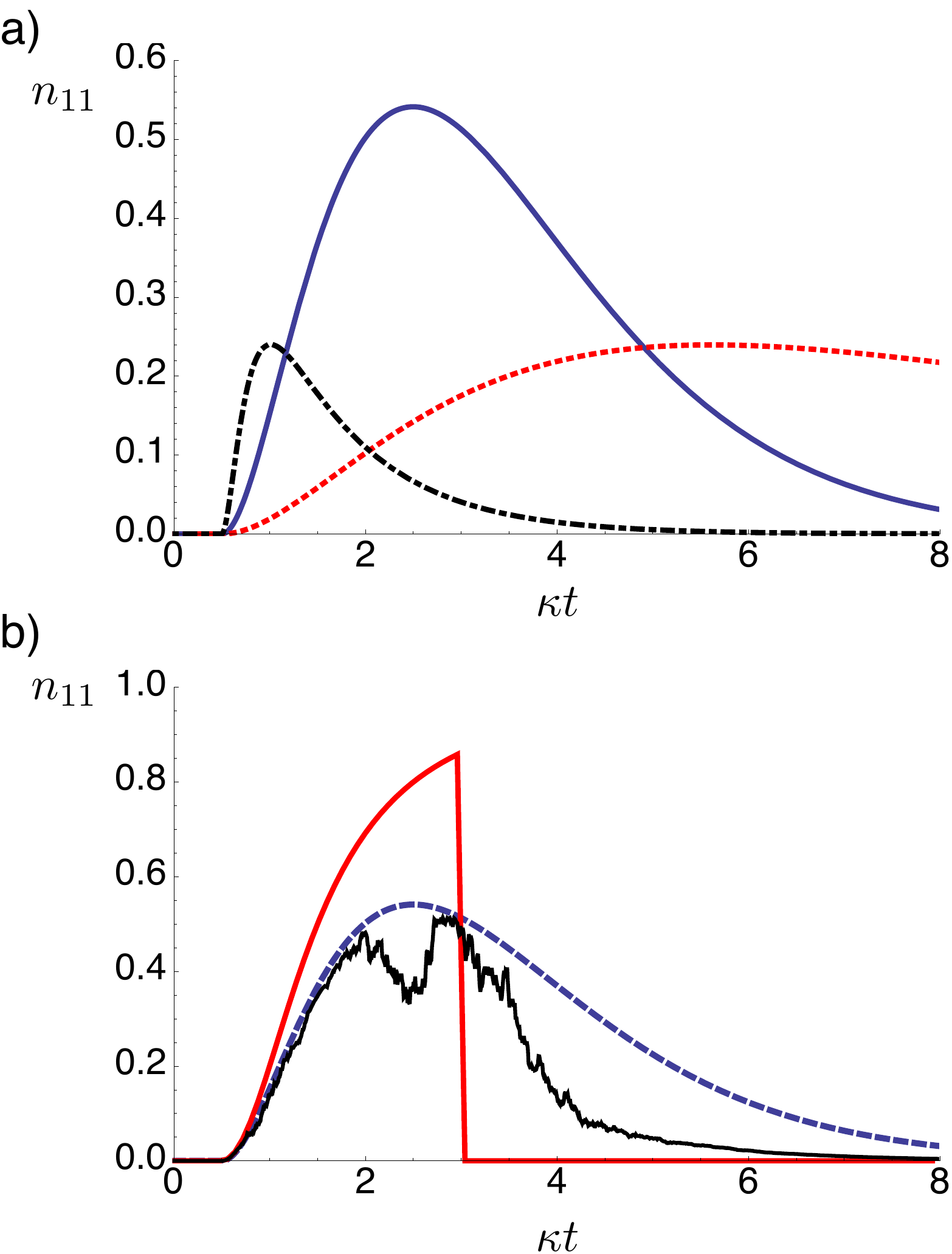}
\caption{Average number of photons in the cavity for a single photon driving. a) Exact master equation solutions for $\gamma/\kappa=1$ (blue, solid line), 10 (black, dot-dashed), and 0.1 (red, dotted). b) Single trajectories for photodetection (top, red curve) and homodyne (solid black line) monitoring with $\gamma/\kappa=1$. The exact master equation solution is also shown (blue dashed line) for comparison.}
\label{fig:avgN}
\end{figure}

\begin{figure}[htb]
\includegraphics[width=8cm]{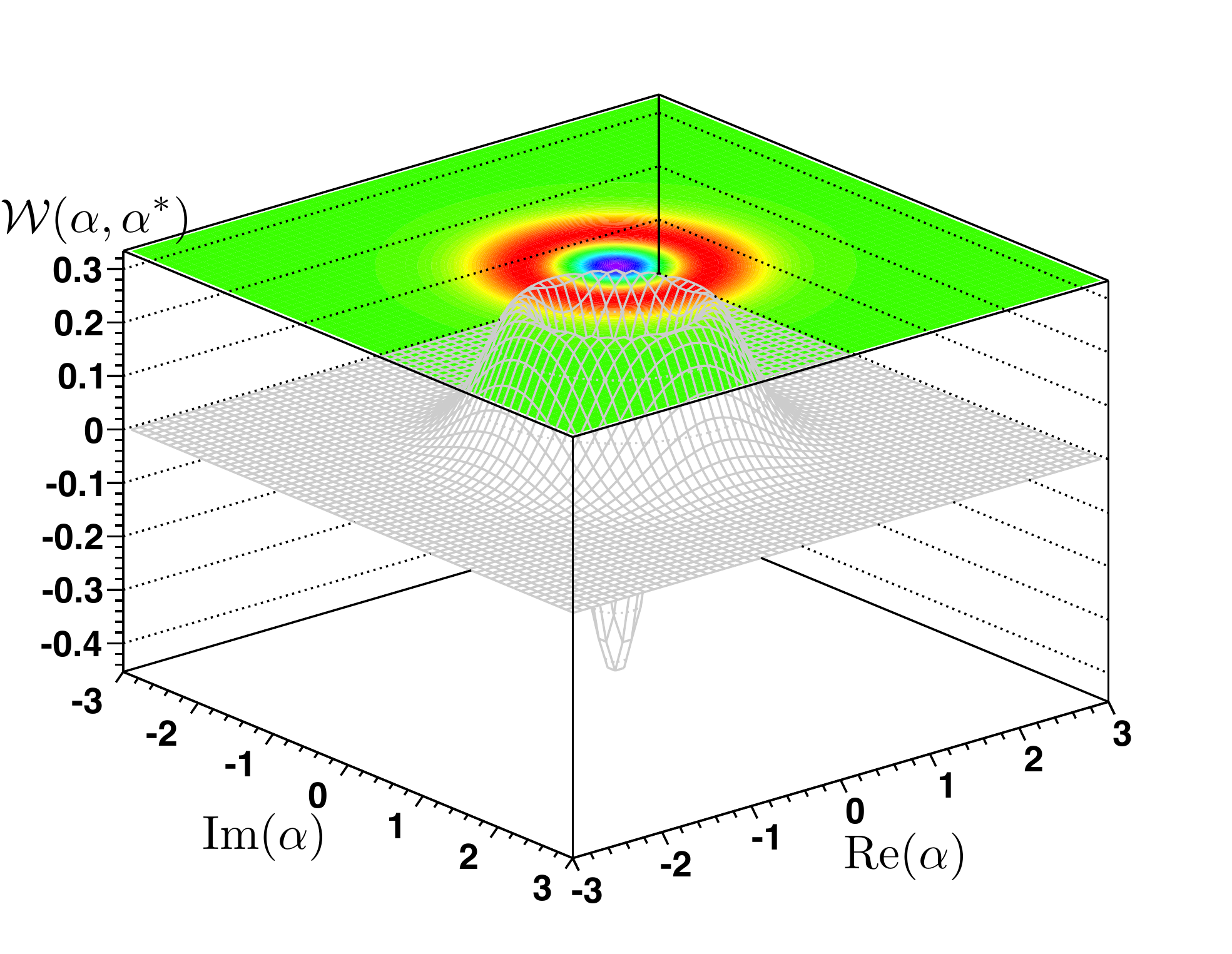}
\caption{Wigner function for the field inside the cavity corresponding to the single photodetection trajectory shown in Fig.~\ref{fig:avgN}-b at $\kappa t=2.8$. Just before the jump the field is in a mixture between the vacuum and a single photon state.}
\label{fig:wigner}
\end{figure}

This difference becomes evident when the system is being monitored. For the photocounting monitoring the set of equations needed to describe the system is 
\begin{subequations}
\label{eq:avgNsp}
\begin{align}
d\langle n\rangle_{11}&=\left[-\kappa \langle n\rangle_{11} -\sqrt{\kappa} \langle a\rangle_{01} \xi^*(t) -\sqrt{\kappa}\langle a\rangle^*_{01}\xi(t) \right] dt\nonumber \\ &+\left[\nu_t^{-1} \langle n\rangle_{00}\vert\xi(t)\vert^2-\langle n\rangle_{11} \right] dN(t),\\
d\langle a\rangle_{01}&=\left[ -\frac{\kappa}{2}\langle a\rangle_{01} -\sqrt{\kappa} \langle {\mathbbm 1} \rangle_{00} \xi(t) \right]dt \nonumber \\&+\left[ \nu_t^{-1} \sqrt{\kappa}\langle n\rangle_{00} \xi(t)  -\langle a\rangle_{01} \right] dN(t) , \\
d\langle {\mathbbm 1} \rangle_{00}&= \left[\nu_t^{-1} \kappa \langle n\rangle_{00}  -\langle {\mathbbm 1} \rangle_{00} \right] dN(t),\\
d\langle n\rangle_{00}&=  -\kappa \langle n\rangle_{00}dt -\langle n\rangle_{00} dN(t).
\end{align}
\end{subequations}
Note that, because we assume that the cavity starts in the vacuum, at any given time there can be at most one photon in the cavity and we neglected expectation values with higher order in $a$ and $a^{\dagger}$ such as $\langle a^{\dagger}a^{\dagger} a\rangle_{ij}$ in Eqs.(\ref{eq:avgNsp}). Note also that $\langle n\rangle_{00}(0)=0$ and only the first three equations need to be considered to examine the conditional number of photons in the cavity. For the coherent driving the equation is
\bea
d\langle n\rangle&=& \left(-\kappa \langle n\rangle +i(\epsilon(t) \langle a\rangle -\epsilon^* \langle a^{\dagger})\right) dt \nonumber \\
&+& \left(-\langle n\rangle + \langle {a^{\dagger}}^2a^2\rangle/\langle n\rangle \right) dN.
\eea
In this case, the stochastic term cancels (the state is an eigenstate of the jump operator $a$), showing that the cavity field is not affected by detection events. The evolution therefore coincides with the unmonitored one and is shown by the dashed line in Fig.\ref{fig:avgN}-b. The conditional dynamics for a single photon driving is quite different:  In the case of a homodyne monitoring the single photon entering the cavity is perturbed by the measurement noise and ends up evolving back to the vacuum state, while for photodetection a detection induces a jump back to the vacuum state (see Fig.\ref{fig:avgN}-b). 

\section{Two-mode cavity and nonlinear phase shifts}
\label{sec:phase}

The results from the previous section will help us analyse the situation shown in Fig.~\ref{fig:2modes}. A doubly resonant cavity containing a Kerr nonlinear medium is driven by a single photon in one mode (mode $a$) and a coherent field in the other (mode $b$). Due to the nonlinear interaction between the modes, the presence of a photon in mode $a$ will induce a phase shift in mode $b$. This nonlinear cross-phase modulation is the basis for the proposal of all-optical two-qubit gates~\cite{Munro:2005} and has been recently analysed from the point of view of multi-mode propagating single photon fields~\cite{Shapiro:2006,Shapiro:2007,Munro:2010}. The Hamiltonian describing the system is 
\bea
\label{eq:ham}
H=\chi b^{\dagger} b \, a^{\dagger} a +\delta_a a^{\dagger} a + (\beta^* b+\beta b^{\dagger}),
\eea
with $\chi$ being the nonlinearity, $\delta_a$ the detuning of mode $a$ and $\beta$ the coherent field driving mode $b$. Here we will examine the dynamics of this system when only the mode driven by the single photon field is monitored. While this will shed some light into the conditional dynamics of the system, our aim here is not to explore in depth the cross phase modulation problem but rather illustrate the use of the quantum filtering equations described in Sec.~\ref{sec:model} in a more complex scenario. 
\begin{figure}[htb]
\includegraphics[width=0.7\linewidth]{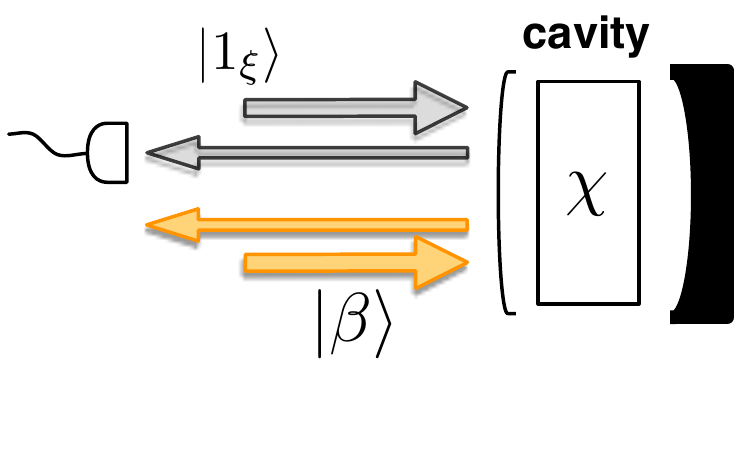}
\caption{Schematic representation of a doubly resonant cavity containing a Kerr nonlinear medium. The cavity is driven by a coherent field in one mode and by a single photon in the other. The output of the single photon mode is monitored by a photodetector while the other mode remains unobserved.}
\label{fig:2modes}
\end{figure}

To describe this system we use the stochastic equations presented in Section~\ref{sec:model}. Note that because only mode $a$ is being monitored, operators for the $b$ mode only appear in the Hamiltonian and in the decoherence term. We then simply need to replace $L=\sqrt{\kappa_a} a$ and ${\cal L} \rho= -i \left[H,\rho\right] + {\cal D}[\sqrt{\kappa_a} a]\rho + {\cal D}[\sqrt{\kappa_b} b]\rho$ in Eq.~(\ref{eq:smepd}) to obtain the exact conditional dynamics under photodetection. The equation for  $\rho_{11}(t)$ (\ref{eq:smepd}a) can be written explicitly as
\bea
\label{eq:rho112modes}
d\rho_{11}(t)&=& \left[ -i \left[H,\rho\right] + {\cal D}[\sqrt{\kappa_a} a]\rho + {\cal D}[\sqrt{\kappa_b} b]\rho\right]dt \nonumber \\
&+&\left[\sqrt{\kappa_a} [\rho_{01}(t),a^{\dagger}] \xi(t) +\sqrt{\kappa_a}[a,\rho_{10}(t)]\xi^*(t)  \right] dt \nonumber \\
&+&\left[ \right. \nu_t^{-1}\left( \right. \kappa_a a\rho_{11}(t)a^{\dagger}  +\sqrt{\kappa_a} a\rho_{10}(t) \xi^*(t) \nonumber \\  &+&\sqrt{\kappa_a} \rho_{01}(t)a^{\dagger}\xi(t) +\rho_{00}\vert\xi(t)\vert^2\left. \right)-\rho_{11}(t) \left. \right] dN(t). \nonumber \\
\eea
The changes in mode $b$ due to the single photon in mode $a$ can then be obtained by writing the equations of motion for the expectation values of the field $b$. Using Eq.~(\ref{eq:rho112modes}) (and the equivalent ones for the other $\rho_{ij}$ components) together with the relation $d \langle b\rangle_{11}=\tr[b\, d \rho_{11}]$, we get
\bea
&&d \langle b\rangle_{11}=\left(-i \beta -i \chi \langle b\, n_a\rangle_{11} -\frac{\kappa_b \langle b\rangle_{11}}{2}\right)dt + \left[-\langle b\rangle_{11} \right. \nonumber \\&& \left.+\nu^{-1}_t\left(\langle b\, n_a\rangle_{11} +\langle b\, a\rangle_{10} \xi^* + \langle b\, a^{\dagger} \rangle_{01} \xi +\langle b\rangle_{00} |\xi|^2\right)\right] dN. \nonumber \\
\eea
Similarly to the single mode case, the equation for $\langle b\rangle_{11}$ couples to other moments and the system of equations can get cumbersome specially in the conditional case. For this reason we omit these equations and will focus now on the numerical simulations of the stochastic equations. 

The introduction of the extra mode $b$ as compared to the situation analysed in Sec.~\ref{sec:cav} increases the dimensionality of the problem and suggests that it could be more efficient to use the SSE (\ref{eq:ssepd}) instead of the SME (\ref{eq:smepd}). However, the model that we propose introduces an extra difficulty which lies in the fact that mode $b$ is not being monitored and therefore the system needs to be described by a density matrix. In this case it is still possible to use a description in terms of pure states via Eq.(\ref{eq:ssepd}) but it is necessary evolve an ensemble of pure states with appropriate weights~\cite{Jacobs:2010}. For the parameters we used, simulations using the full SME turned out to be more efficient due to the excessive number of paths needed for convergence using the SSE.

For our simulations we fixed the decay rate of mode $a$ and set $\gamma=\kappa_a$ so that we have the best matching between the cavity and the incoming single photon as explained in Sec.~\ref{sec:cav}. We also assume that, as the system is monitored, we estimate the number of photons in mode $b$ and feedback this information to adjust mode $a$ detuning as $\delta_a(t)=-\chi \langle n_b(t)\rangle$. This ensures that mode $a$ is always resonant even with the nonlinearity present, as can be seen from the unconditional equation for $\langle a\rangle_{11}$
\begin{equation}
\frac{d \langle a\rangle_{11}}{dt}=-i \delta_a \langle a\rangle_{11}-i \chi \langle n_b\, a\rangle_{11} -\frac{\kappa_a \langle a\rangle_{11}}{2} -\sqrt{\kappa_a} \xi(t) \langle {\mathbbm 1}\rangle _{01}.
\end{equation}
The phase shift imprinted in mode $b$ is proportional to the product of the nonlinearity $\chi$ and the amplitude of the coherent field $\beta$. For realistic cavity parameters the nonlinearity is small so a large $\beta$ is needed for the change in the $b$ field to be appreciable. To be able to see the effects and yet have simple simulations, we fixed the nonlinearity at a relatively large value $\chi=\kappa_a/10$ so that the basis describing mode $b$ could be truncated at a numerically reasonable level. 

Fig.~\ref{fig:2modesresults} shows the $X_b={\rm Re}(\langle b\rangle_{11})$ quadrature of mode $b$ as a function of time for different values of $\kappa_b$. This quadrature is representative of the change in mode $b$ as the orthogonal quadrature $P_b$ doesn't change appreciably during the process. For each value of $\kappa_b$ we adjusted the coherent driving to keep constant the ratio $\Delta \beta = 4\beta \chi/\kappa_b^2$, which corresponds to the maximum conditional change in mode $b$ found in~\cite{Munro:2010} for $\kappa_b \gg \chi$. In their model, Munro {\it et al.} calculate the phase shift assuming that at some random time there will be a single photon in mode $a$ ($n_a=1$) and that it will remain there for some random interval $T$, after which the photon leaves the cavity. Because we can't possibly know when the photon enters mode $a$, in our description we only use the information about the single photon shape and the detection clicks at the cavity output. 

\begin{figure}[htb]
\includegraphics[width=0.9\linewidth]{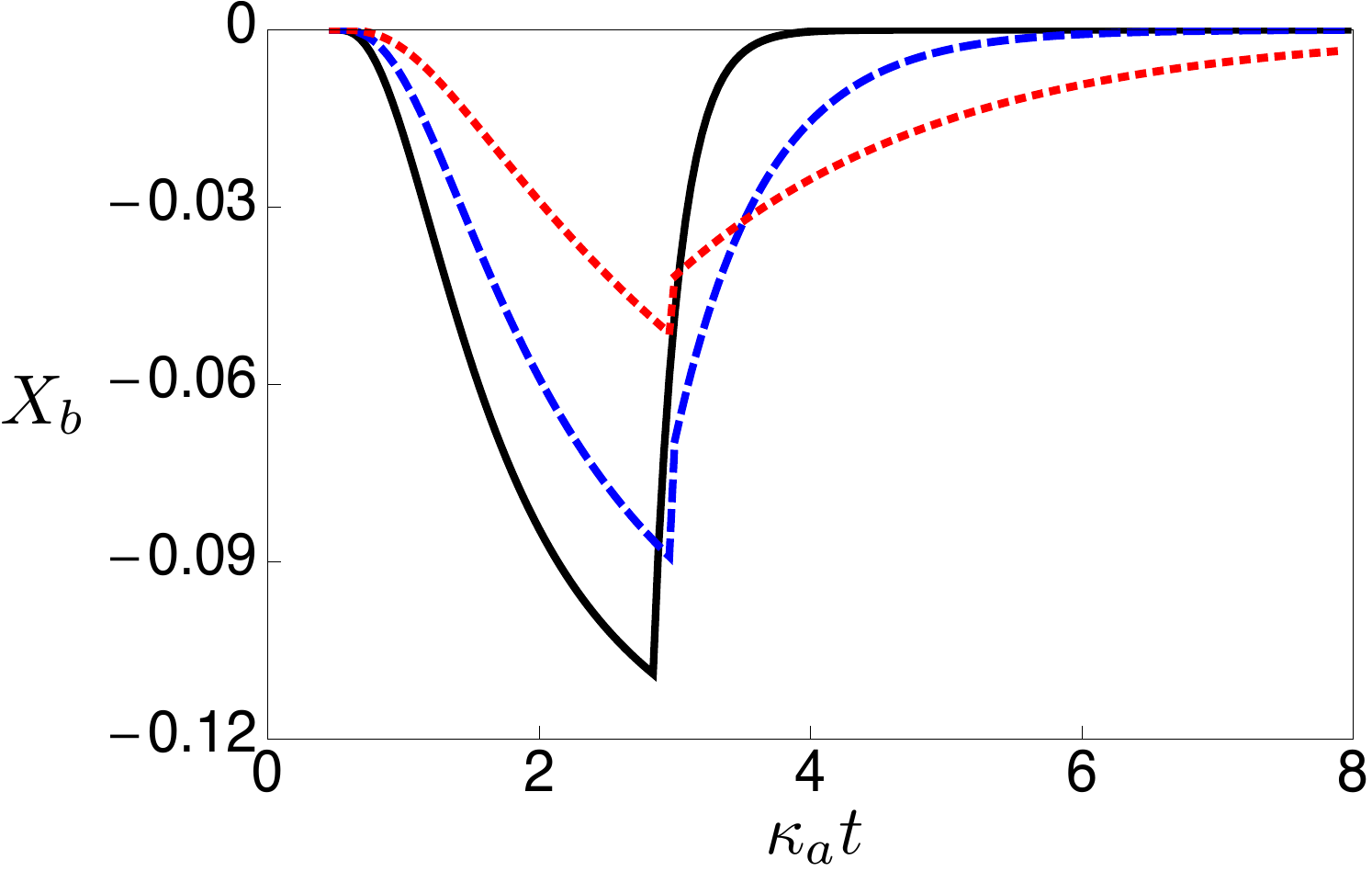}
\caption{Evolution of the $X$ quadrature of mode $b$ for a single trajectory with $\kappa_b/\kappa_a=10$ (solid, black), $3$ (dashed, blue) and $1$ (dotted, red). The coherent driving is set to $\beta=\kappa_b^2/(4 \kappa_a)$ such that $\Delta \beta=0.1$.}
\label{fig:2modesresults}
\end{figure}

We found that by increasing $\kappa_b$ not only the time taken for mode $b$ to relax back to steady state after a detection event decreases, but also the maximum phase shift increases. This is in qualitative agreement with the results in~\cite{Munro:2010} which shows that the maximum phase shift is obtained for $\kappa_b \gg \kappa_a$. However, the conditional phase shifts obtained here can exceed the maximum possible value, $\Delta \beta$, found in~\cite{Munro:2010}.  

To explore how our exact stochastic simulations can bring new information about the conditional phase shift process, we calculated the maximum conditional phase shift for a set of $5000$ trajectories and constructed the histogram depicted in Fig.~\ref{fig:histo} for (a) $\kappa_b=\kappa_a/2$ and (b) $\kappa_b=4 \kappa_a$. As the ratio $\kappa_b/\kappa_a$ gets larger, the change in the $b$ field gets sharper (see Fig.~\ref{fig:2modesresults}) and this is mapped to a distribution that favours larger phase shifts as shown in Fig.~\ref{fig:histo}-b. An interesting aspect that our model can capture is the structure of the distributions in Fig.~\ref{fig:histo}. They are divided in 2 parts: one corresponding to small phase shifts and therefore to detection events that occurred shortly after the photon in mode $a$ enters the cavity, and another one with larger shifts representing detections after mode $a$ had enough time to build up in the cavity. The two parts are divided by a gap region with values that do not show up in any trajectory. This region can be understood by looking at the probability of detecting a photon between $t$ and $t+dt$, which is given by $P_J=\nu_t \, dt$ with 
\begin{equation}
\label{eq:nu}
\nu_t =\kappa_a \langle n\rangle_{11} +\sqrt{\kappa_a} \left(\langle a \rangle_{01}\xi^*(t) +\langle a\rangle_{01}^* \xi(t) \right)+\langle {\mathbbm {1}}\rangle_{00} |\xi(t)|^2.
\end{equation}
For a cavity that is not driven, the rate of detection is given by the first term in Eq.(\ref{eq:nu}), {\it i.e.} it is simply proportional to the decay rate and the mean number of photons in the cavity. When the driving is on, all the terms have to be considered to account for the interference that leads to the dip in $\nu_t$ shown in the inset of Fig.~\ref{fig:histo}. At that time no photons are expected to come out of the cavity as $\nu_t$ goes to zero. Note that this behaviour is independent of the coupling between the two modes and appears also in the single mode problem discussed in Section~\ref{sec:cav}. A simple calculation of the field output intensity also shows that this effect occurs for a coherent driving and does not rely on the single photon character of the field.
\begin{figure}[htb]
\includegraphics[width=0.95\linewidth]{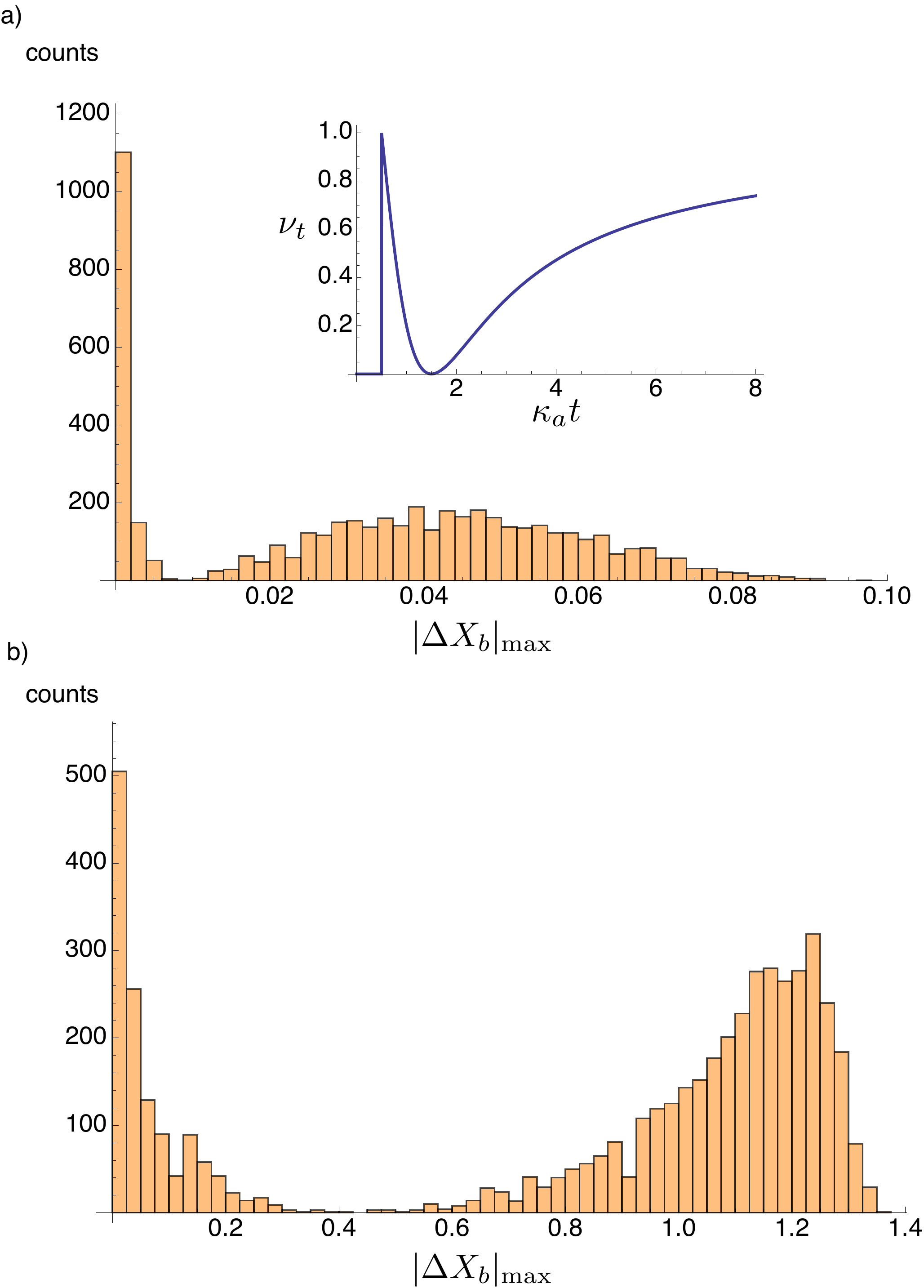}
\caption{Histogram of the maximum conditional change in the $b$ mode for $5000$ different trajectories and (a) $\kappa_b=\kappa_a/2$ and (b) $\kappa_b=4 \kappa_a$. The inset shows $\nu_t$ for a particular trajectory where no photons have been observed up to the time shown in the plot. The gaps in the distributions exists because there are no photons coming out of the cavity when $\nu_t=0$.}
\label{fig:histo}
\end{figure}

\section{Conclusions}
\label{sec:conc}
We have presented the stochastic Sch\"odinger equations corresponding to the quantum filter, or stochastic master equations, for systems driven by single photons. We applied this trajectory formalism to investigate the conditional phase shift of a field that interacts nonlinearly with a single photon pulse. By presenting a complete quantum stochastic model and  numerically solving it, we corroborated some of the results in Ref.~\cite{Munro:2010} but were also able to predict different effects on the distribution of conditional phase shifts. 
The formalism applied here is well suited to model measurement and feedback and we expect that the approach will be useful to describe other systems such as quantum logical operations, quantum repeaters, or the interaction of nodes in a quantum network connected by single photons.

\acknowledgments{We thank M. Hedges, J. Combes, J. Gough and H. Nurdin for useful discussions. This research was conducted by the Australian Research Council Centre of Excellence for Quantum Computation and Communication Technology (Project number CE110001027). We also thank the support from AFOSR Grant FA2386-09-1-4089 AOARD 094089.}

\bibliography{$HOME/ARTICLES/allbib}

\end{document}